\documentclass[12pt]{article}

 \usepackage{amsfonts}
 \usepackage{amssymb,amsmath}
 \usepackage{graphicx} \usepackage{epstopdf}
 \newcommand{\crlb}[1]{\label{#1}\\[2pt]}
 
 \newcommand{\crld}[1]{\label{#1}}
 \newcommand{\eela}[1]{\quad\hbox{\scriptsize{#1}}\label{#1}\end{eqnarray}}
 \newcommand{\eelb}[1]{\label{#1}\end{eqnarray}}
 
 \newcommand{\newsecb}[2]{\section{#1}\label{#2}\setcounter{equation}{0}}

 \newcommand{\nolabels} {\def\eel{\eelb}\def\eeql{\eeqlb}  \def\crl{\crlb} \def\newsecl{\newsecb}\def\bibiteml{\bibitem} \def\citel{\cite}\def\labell{\crld}}
\newcommand{\eeqla}[1]{\quad\hbox{\scriptsize{#1}}\label{#1}\end{aligned}\end{equation}}
\newcommand{\eeqlb}[1]{\label{#1}\end{aligned}\end{equation}}

\newcommand\publishversion{\nolabels\setlength{\textheight}{8.3in}\setlength{\oddsidemargin}{0in}
   	 \setlength{\textwidth}{6.3in}\setlength{\topmargin}{-0.2in}}

\def\beq{\begin{equation}\begin{aligned}}		\def\eeq{\end{aligned}\end{equation}}
\def\be{\begin{eqnarray}}  					\def\ee{\end{eqnarray}}		
\def\bi#1{\begin{itemize}\item[#1]} 			 			\def\ei{\end{itemize}} 
  \def\eqn#1{(\ref{#1})}
   	 \def\fn{\footnote}	  		 

		 \def\a{\alpha}   \def\b{\beta}           \def\l{\lambda}  

    		  	   \def\e{\varepsilon} 
	    		\def\f{\phi}        		     		     
 	 		\def\s{\sigma}     	      	 
	     		          		              	  
    		  		\def\dd{{\rm d}}

 		\def\ket{\rangle}

\def\fract#1#2{{\textstyle\frac{#1}{#2}}}	 	 	
\def\ffract#1#2{\raise .2 em\hbox{$\scriptstyle#1$}\kern-.3em/\kern-.2em\lower .15 em \hbox{$\scriptstyle#2$}}

\def\bpmatrix{\begin{pmatrix}} 			\def\epmatrix{\end{pmatrix}}
\def\bmatrix{\begin{matrix}} 			\def\ematrix{\end{matrix}} 
\def\bcenter{\begin{center}}			\def\ecenter{\end{center}}

\def\weglaten#1{}

\def\lowerheightfig#1#2#3{\(\raise-#1\hbox{\includegraphics[height=#2]{#3}}\)}
\def\lowerwidthfig#1#2#3{\(\raise-#1\hbox{\includegraphics[width=#2]{#3}}\)}
\def\th{\({}^{\mathrm{th\,}}\)}		 
   
\def\ini{{\mathrm{init}}}  \def\fin{{\mathrm{final}}}  \def\ont{{\mathrm{ont}}}

\publishversion

\begin{document}

\begin{titlepage}
 \title{\vskip 20mm \LARGE\bf Time, the arrow of time, and Quantum Mechanics}
\author{Gerard 't~Hooft}
\date{\small  Institute for Theoretical Physics \\ Utrecht University  \\[10pt]
 Postbox 80.089 \\ 3508 TB Utrecht, the Netherlands  \\[10pt]
e-mail:  g.thooft@uu.nl \\ internet: 
http://www.staff.science.uu.nl/\~{}hooft101/}
 \maketitle

\begin{center} Submitted to \emph{The Quantum and the Classical: \\
Emergence, Computation and the Origin of Time}
\end{center}

\vfill

\noindent \textbf{Abstract} \\[-25pt]
\begin{quotation}
\noindent It is brought forward that viable theories of the physical world that have no variable at all that can play the role of time, do not exist; some notion of time is one of the very first ingredients a candidate theory should possess. Almost by definition, time has an arrow. In contrast, time \emph{reversibility}, or even the possibility to run the equations of motion backwards in time, is not at all a primary requirement. This means that the direction of the arrow of time may well be uniquely defined in the theory, even locally. We explain these statements in terms of the author's favoured deterministic cellular automaton interpretation of quantum mechanics, also to be referred to as `vector space analysis', and expand on these ideas.\end{quotation}
\end{titlepage}
\newsecl{Introduction; defining time}{intro}
	The universe as we know it is characterised by a framework called \emph{space-time}, in which \emph{events} take place. The events are characterised first of all by their \emph{locations} in space, and \emph{moments} in time, all together indicated in terms of coordinates. The number of coordinates needed, usually real numbers,   is called the dimension of space-time. 	 
The coordinate that indicates  time, is a very special one. It is the only coordinate in which it is meaningful to define an ordering in the values given, the order of time. This ordering defines an orientation, called the arrow of time. It allows us to define an ordering (or at least a partial ordering) of all events.

	Whenever we build \emph{models} that explain the existence and nature of the events, it is of extreme importance to have such an ordering of events; it allows us to explain them sequentially: one event can be the cause of a subsequent event if its time variable is lower, or it could be the consequence of an event if its time coordinate is higher. It is difficult, probably impossible, to devise a model of our universe, if no ordering is defined for the model to describe the events. 
	
	This in fact will bring us to provide a definition of time that is more primary, more basic than all other ingredients of our model, including the notion of space. Our universe is known to carry a memory of things that happened in the past. Whenever we build a model of our universe, one that is controlled by ``laws of physics", it should come with a completely unambiguous prescription of \emph{the order in which the laws of Nature should be imposed on all events that take place}. Regard the laws of physics as a computer program to calculate the next sequence of events. The data that we have to enter into the program may come from events calculated earlier. They may not come from events that still have to be specified, because in that case conflicts may arise: if event \(A\) affects the features of event \(B\) then event \(B\) should not react back to modify event \(A\), otherwise the rules cease to be unique; they will literally be circular, making them either self-contradictory or ambiguous, and for that reason they would not be suitable to explain observed phenomena.
	
	Notice that this is the extreme opposite of Newton's action principle: if event \(A\) acts on event \(B\) with some force, event \(B\) should \emph{not} react back onto \(A\). Newton's action principle, action \(=\) reaction, is different because it is in space-like directions, and because it often neglects some minute time delays that are involved: the (re-) action cannot spread faster than the speed of light.
	
	The ordering caused by the rule `\(A\) affects \(B\) but \(B\) cannot affect \(A\)', is one we cannot do without. Assuming indeed that the universe allows for te existence of such an elementary action \(\ne\) reaction principle, we obtain a unique \emph{definition} of time:\\[5pt]
\indent \emph{Time is the order in which our models for nature predict, prescribe or explain events.} \\[5pt] 
	Notice that this definition of time does suppose that we construct models to explain our universe. If one \emph{only} would collect data, without attempting to explain them, we would not need any notion  of time.
	
	Notice also that quantum mechanics provides no exception to our rule; it also requires a definition of an ordered time coordinate. We can say this because the Schr\"odinger equation\fn{Here, and in what follows later, all equations of the form \(\fract{\dd}{\dd t}|\psi\ket=-iH\,|\psi\ket\), where \(H\) is a hermitian operator,  are referred to as Schr\"odinger equation, regardless whether they act on wave functions of more general vectors in Hilbert space.\label{noteschr}} involves exclusively a first order derivative in time. Therefore only one boundary condition is needed, taken to be the situation in some distant past, to determine the situation in the future.

The primary definition of time given above, only defines the time ordering, but does not attach real numbers to time. In fact, the use of integers, so as to count the events that we calculated, would have been more appropriate.
Considering the humongous size of our universe, and the extremely short time sequences expected to be relevant at the Planck scale, one may expect these integers, if they exist at all, to be extremely large, larger than \(\sim 10^{60}\). Scaling these numbers down for practical use, probably suffices to explain why, at present, real numbers seem to be more useful than integers to indicate time.

According to special relativity, one can have events that are \emph{space-like} separated. This means that  there may be events \(A\) and \(B\) such that our model allows us to calculate what happens in \(A\) and in \(B\) without the need to specify their order. The importance of this is that the definition of time given above is not unique; it is a feature of the notion of time that will have to be taken into consideration when building more advanced models, but it seems to be less basic as far as first principles is concerned.

\newsecl{Quantum Mechanics}{QM}
	The theory of quantum mechanics is arguably one of the greatest discoveries of physics; it revolutionised our understanding of molecules, atoms, radiation, and the world of the sub-atomic particles. Yet even now, almost hundred years later, there is still no complete consensus as to what the theory tells us about \emph{reality}, or even whether `reality' exists at all. Some authors adhere to the idea that \emph{all} `realities' exist somewhere in some alternative universes, and that these universes evolve together as  a `multiverse'.\fn{`Multiverse' can mean different things. In cosmology, it means that there may be different regions of our universe where the inflation rates and perhaps also the effective laws of nature, vary. In quantum mechanics, one might view the `many worlds' together as a multiverse.}

\setcounter{page}{2}
The present author does not go along with such ideas. Quantum mechanics is a superb description of the world of tiny things, but, on the face of it, quantum mechanics seems merely to reflect humanity's ignorance. \emph{We do not know} which reality it describes, and as long as this is the case, we should not be surprised that, in a sense, all possible realities play a role whenever we try to make the best possible prediction of the outcome of an experiment. The fact that many of us have technical difficulties implementing such a thought in the equations known to work best today, may well be due to lack of imagination as to how eventually the correct view will be found to emerge.

The author has made his own analysis of the known facts, and came to the conclusion that the Copenhagen doctrine, that is, the consensus reached by many of the world experts at the beginning of the 20\th century, partly during their numerous gatherings in the Danish capital, has it almost right: there is a wave function, or rather, something we call a quantum state, being a vector in Hilbert space, which obeys a Schr\"odinger equation.\fn{See footnote \ref{noteschr}.}  The absolute squares of the vector components may be used to describe probabilities whenever we wish to predict or explain something. Powerful techniques were developed, enabling one to guess the right Schr\"odinger equation if one knows how things evolve classically, that is,  in the old theories where quantum mechanics had not yet been incorporated. It all works magnificently well. According to Copenhagen, however, there is one question one should \emph{not} ask: ``What does \emph{reality} look like of whatever moves around in our experimental settings?", or: what is really going on?

According to Copenhagen, Such a question can never be addressed by means of any experiment, so it has no answer within the set of logical statements we can make of the world. Period, schluss, fini. Those questions are senseless.

It is this answer that we dispute. Even if this kind of questions cannot be answered by \emph{experiments,} we can still \emph{in theory} try to build credible models of reality. Imagine the famous detective Sherlock Holmes entering a room, with a dead body lying on the floor. The door is open, and so is the window. A crime has been committed. Did the perpetrator come through the window or through the door? Or did something altogether different happen? Sherlock Holmes ponders about all possibilities, but he will \emph{not} say: the perpetrator came through the window \emph{and} through the door, using a wave function, etc. etc. Clearly such answers are not accepted in the ordinary world. Sherlock Holmes may well conclude that he cannot derive the answer with certainty, but what he can try to find out is
 what \emph{could have} happened. Have we been brainwashed to accept wave functions in the world of the atoms? Should we not, here also, ask what it really was, or what it could have been, that has been going on? 

Perhaps we are using the wrong language. Maybe atoms and molecules do not exist in the form we imagine them. Maybe Nature's true degrees of freedom are very different, and only when we consider the statistics of many atoms, our language that assumes these to be particles obeying quantum equations may be seen to work out correctly.

When early attempts to construct such models failed, investigators tried another path: maybe one can \emph{prove} that there exists no reality at all whose probabilities can be caught in terms of a Schr\"odinger equation? Suppose that we impose conditions on such models such as \emph{locality} and \emph{causality}. Can one prove or disprove that realistic models exist?

What then happened is well-known. The first to consider such an option was Einstein, together with his co-authors Podolsky and 
Rosen\,\cite{EPR}. They conceived of a \emph{Gedanken} experiment to show that quantum mechanics cannot exactly provide a local description of what is going on. 
This conclusion is in fact somewhat contradictory, because quantum mechanics \emph{was} used to describe as accurately as possible what predictions can be made, and that result was rarely disputed by anyone; indeed it was confirmed later by real experiments.

The setup was revised by a somewhat more realistic scenario using particle spins, by J.\,S.~Bell, and he gave the apparent contradiction in a more precise wording: Bell's theorem: \begin{quote}
\emph{No physical theory of local hidden variables can ever reproduce all of the predictions of quantum mechanics;}\end{quote}
the outcome of a quantum mechanical calculations of some non-local correlations contradicts any acceptable `classical' explanation by at least a factor \(\sqrt 2\). The inequality, called \emph{Bell inequality}, was subsequently generalised and made more precise\,\cite{CHSH}.
 
 \newsecl{Causality, correlations and quantum mechanics}{causal} 
 This finding did not go undisputed. Many authors attempted to locate the flaw in Einstein's and Bell's argument, but logically it seemed to be impeccable. Bell assumed that determinism means that one can build a model, any model, in which classical equations control the behaviour of dynamical variables, and where, at the tiniest scales where these variables describe the data, the evolution laws do not leave the slightest ambiguity; there are no wave functions, no statistical considerations, as everything that happens is controlled by certainties. Moreover, there is some sense of locality: the laws control all processes using only the data that are situated at given localities, while action at a distance, or backwards in time, are forbidden. The classical degrees of freedom that `really' exist were called `beables'.
 
Here, the first topic for discussion arises: \emph{what does `action backwards in time' mean?} In ``La Nouvelle Cuisine", Bell\,\cite{Bell}  formulated as precisely as he could what `causality forward in time' means: \begin{quote} A theory is said to be locally causal if the probabilities attached to values of local beables in a space-time region 1 are unaltered by specification of values of local beables in a space-like separated region 2, when what happens in the backward light cone of 1 is already sufficiently specified [...] \end{quote}
Region 2 is assumed to be completely outside the past light cone of 1, so what happens there, must be immaterial. 
It sounds fine, and many researchers agree with it, but there is a problem:\begin{quote}
Region 2 also has a past light cone, and if we consider some modification of the events in 2, these may disagree with what we postulated to have in region~1, since the two past light cones overlap. \end{quote} Consequently, \emph{correlations} between the data in region 1 and region 2 \emph{cannot} be excluded. In fact, such correlations are known to occur ubiquitously in the physical world, so what does ``Bell-causality" really mean?

What Bell needed to have said is that, \emph{in any model} describing the laws of nature, only the data in the past light cone of 1, should determine what happens in 1, while he should not have referred to correlations. \emph{Yet Bell's inequalities are about correlations}, and these are assumed to be absent outside the light cone.

In the same vein, `backwards causality' is rejected: the past should not depend on the future. This is true in the following sense: \emph{our model} should not require knowledge of the data in the future, to prescribe the data at present (it should only require data in the past light cone). Correlations do occur. In fact, if our model reflects reversibility in time -- which most models do -- then the data inside the future light cone \emph{can be used} to determine, that is, to reconstruct,  the present or the past, back from the future. 

In the above, the words \emph{our model} were emphasised. What is important here is that causality cannot be a feature or property of the physical data themselves, but rather a property of the equations of motion with which we try to mimic these data. If two different theories can be used to describe the same set of data, then one of these theories might have causality and the other not. This is an element of the Bell ``paradox" that may not have been emphasised sufficiently.

Most models of nature are reversible in time; we can run the basic equations backwards in time as easily as forwards in time\fn{We refer here to the equations at small time intervals and accordingly acting at small distance scales. Thermodynamics on the other hand, valid for large time intervals cannot be easily inverted in time.}. This implies that theories with causality forwards in time must also have causality backwards in time; this was ignored by Bell.

There is nevertheless a good reason why Bell's profound result is considered irrefutable by most researchers today.  The actions of observers in quantum experiments, are considered to be completely classical, and they reflect the observers' \emph{free will}. To overrule Bell's theorem, the observers' free will must be correlated with quantum data in the past. This is considered `absurd' by most researchers. In the next section, and in Appendices \ref{Born} and \ref{6bit}, this author's response, as to why these correlations may be not so absurd after all, is further illuminated.

The theory used by the author was called ``Cellular Automaton (CA) Interpretation"\,\cite{GtHCA}, but perhaps a preferable denomination is ``vector space analysis\fn{The phrase `vector space analysis' is used in information technology; it is the same mathematics that is used there. We add to that procedures involving unitary transformations.}\,". It is the idea that a classical system may be analysed by associating any state of the system by a vector, such that all states together form an orthonormal basis of a vector space called Hilbert space. ``vector space analysis" consists of the mathematical procedures made possible by performing any kind of transformations in this vector space. One ends up with a Schr\"odinger equation exactly as in quantum mechanics. Thus, vector space analysis contradicts Bell's theorem. Our theory consists of the assertion that what we call quantum mechanics today can be the result of a vector space analysis of some classical system. The ``CA Interpretation" of quantum mechanics consists of the assumption that this is true, while we refrain from further attempts to identify the classical system underlying it. The author hopes however that the search for appropriate classical models will continue, and that it will bear fruit.

We end this section with the remark that a restriction exists called ``causality", that can be imposed on any model for elementary particles. It is not disputed, but in fact used a lot in quantised field theories. This condition considers operators \(\f(x)\) in quantum field theories, describing (elementary or composite) fields \(\f\) at 4-space-time coordinates \(x\). Let \(x_1\) and \(x_2\) be space-like separated: \((x_1-x_2)^2>0\). Then we have for the commutator,
\be [\f(x_1),\,\f(x_2)]=0\ .\eel{spacelikecomm}
This says that any operation \(\f(x_1)\) acting on any quantum state at space-time point \(x_1\), cannot affect the result of any dynamical effect \(\f(x_2)\) occurring at \(x_2\). In the Standard Model for the elementary particles, this condition, ``no Bell telephone", is found to hold true, and it has important applications in calculations. However, this condition does not distinguish causal relations in the forward time direction from ones in the backwards time direction, so that it could not be used to derive inequalities such as Bell's.

\newsecl{The Bell and CHSH inequalities}{obs}

Bell's Gedanken experiment is in essence much the same as the Einstein Rosen Podolsky set-up. A local device is constructed that can emit two entangled particles, \(\a\) and \(\b\), which leave the machine in opposite directions.  Alice (\(A\)) and  Bob (\(B\)), both choose whether to measure property \(X\) or property \(Y\) of the particles they can see. Alice chooses setting \(a\) to measure \(\a\) and Bob setting \(b\) to measure \(\b\).

The correlations needed to explain the quantum mechanical result require that the settings \(a\) and \(b\) chosen by Alice and bob, must be correlated with one another as well as the (classical) spins of the two entangled particles. The author calculated the minimal amount of correlation that is needed to produce the quantum result. We found the following distribution\,\cite{GtHCA}:
\be W(a,b,\,\l)=C|\sin(2a+2b-4\l)|\ , \eel{mousedroppings}
where \(a\) is the angle chosen by Alice for her measurement, \(b\) is Bob's angle, and \(\l\) a parameter describing the polarisation of the entangled photons produced by the source -- and detected by Alice and Bob. \(W\) is the probability distribution, and \(C\) is a normalisation constant. It features a \emph{3 body correlation}: whenever we integrate over all values of \(a\), or all values of \(b\), or all values of \(\l\), we get a flat distribution.

To show rigorously that such correlation features are unacceptable for any theory that generates quantum mechanics from classical mechanical laws, Bell had to formulate his definition of causality. We indicated above that his definition does not apply for physical systems, so one could terminate the discussion here and now, since correlation functions are not bounded by light cones. Yet the correlation function \eqn{mousedroppings} is considered unacceptable by most investigators. How can it be that decisions by Alice and Bob, made out of free will, can yet be correlated with something that happened earlier -- the polarisation chosen by the entangled photons emitted by the source? Did these photons ``know" what settings Alice and Bob would later choose, or is this a case of `conspiracy'? How can a single photon guide the classical dynamical variables \(a\) and \(b\)?

To explain this, we now summarise how vector space analysis works.
Suppose we have a classical theory at, for instance, the Planck scale, \(10^{-33}\) cm. This would be typically a cellular automaton. It can be in \(2^{10^{99}}\) states in every \(\mathrm{cm}^3\), typically. Every one of these states is called `ontological', which means it is realised or it is not realised, but superpositions do not exist. It is precisely the thing that Einstein, Bell and others wanted to disprove. Just in order to do mathematics, we now attach a basis vector to every one of these ontological states. They are set up such that they form an orthonormal basis of a \(2^{10^{99}}\)-\,dimensional vector space, at each \(\mathrm{cm}^3\). At the beat of a clock, typically with the Planck frequency of some \(10^{44}\) Hertz, these states evolve into other states. This we write using the evolution matrix, which consists of one 1 in each row and in each column, and zeros everywhere else.

The math we use consists of diagonalising this matrix. This gives us the eigen states of the energy, \emph{i.e.} the Hamiltonian. One finds that the states of this model obey the Schr\"odinger equation. 
Now all energy eigen states are superpositions of the ontological states, and if we limit ourselves to states with energies below 1 TeV for every excitation, then this corresponds to a very tiny subspace of the entire Hilbert space, while every state we can use is a superposition of ontological states. Without loss of generality, we can interpret the coefficients of these superpositions by taking their absolute squares to indicate probabilities. This is further elucidated in Appendix \ref{Born}. Here it is important to observe that `reality' is always described as one of te original ontological states, and never a superposition, yet we may use the Schr\"odinger equation to describe both the ontological states and the superpositions. The elements of the ontological basis always evolve into other elements of this basis, and superpositions into superpositions. We call this the \emph{law of conservation of ontology}.

There is a good reason why many attempts at making realistic models explaining the violation of Bell's inequalities failed, which is that, in these models, it was attempted to mimic superpositions of particular modes in terms of other valid modes of an automaton.  It is much better to keep superpositions as what they are, superpositions of valid automaton modes which, for that reason cannot by themselves act as ontological states. What happens instead is that, if one considers some superposition of physical states, one is actually considering a probabilistic mixture, but what exactly the true, unmixed, physical states are differs from one experiment to the next, in such a way that the final state can never be in a superposition. Because this feature is of tremendous importance, we explain some technical details of this point in Appendix \ref{Born}.

Now we can see that, in deriving their inequalities, Bell and CHSH had to make assumptions that we cannot agree with.
Their main assumption is that Alice and Bob may choose what to measure, and that this should not be correlated with the ontological state of the entangled particles emitted by the source. However, when, in choosing their settings,  either Alice or Bob change their minds ever so slightly, their classical settings represent a different ontological state than before. The photon they look at now, will be a superposition of the old photons that they wanted to detect, but the entire state, photon plus settings, will be orthogonal to the previous one. In particular, because of the ontological conservation law, the new photon they look at must be an ontological one. Alice and Bob do \emph{not have the free will} to look at photons that are not ontological. So, while changing their minds, Alice and/or Bob had to put the universe in a different ontological state than the previous state, and this modification goes back billions of years, all the way to the origin of the universe. One could call this retro-causality, but it is merely due to the fact that the (classical as well as quantum) equations can, in principle,  be solved backwards in time.

As a consequence,  Alice's and Bob's settings can and will be correlated with the state of the particles emitted by the source, simply because these three variables do have variables in their past light cones in common. The change needed to realise a universe with the new settings, must also imply changes in the overlapping regions of these three past light cones. This is because the universe forces itself to stay ontological at all times.

The restriction that the universe must be in an ontological state at all times, is the only restriction. This implies that Alice and Bob still have free will in the classical sense; they can choose any of the ontological states of the universe, no matter what kind of random number generator or lotto machine they were using. But they cannot put the universe in a superposition of states, which is only something we can do in our mathematical models when studying probability distributions,  wishing to bring these in a form such that we can apply Schr\"odinger equations.

A  related quantum paradox that has been put forward as another illustration of quantum weirdness, is the so-called GHZ paradox. This paradox is of interest because its resolution can be phrased in terms of an over simplified model of the universe, illustrating the important role of the  \emph{observer} as being part of the system. In appendix \ref{6bit}, we explain what happens in the cellular automaton theory when this Gedanken experiment is performed.

\newsecl{Information Loss and the Arrow of Time}{infoloss}
Most well-known  physical theories that explain the apparent absence of time reversal symmetry contain elements of thermodynamics and entropy. Actually, in these descriptions of nature, one can explain the absence of this symmetry elegantly by blaming it to an asymmetry in the \emph{boundary conditions}. When writing differential equations for the laws of nature, one always has to add what we know about the boundaries. As for the boundaries in the space-like directions, little is known, since the universe looks very homogeneous, and no boundary effects have ever been detected. The universe is either strictly infinite in the space-like directions, or we live on a spatially compact manifold such as a 3-sphere or a torus. These boundary conditions show much symmetry.

In the time-like direction, however, there cannot be complete symmetry. The universe appears to have started extremely small, conceivably it all started in a single point. That point must have been highly ordered, having total entropy very small or possibly zero. This is a reasonable boundary condition at the origin of time.

Yet at the other end, when time grows to be very large, we see no need of any boundary condition; the universe may simply continue to expand forever, undergoing perpetual increase of entropy. Thus we have equations that are symmetric under time reversal but asymmetric in their boundary conditions. This suffices to explain the time asymmetry we see today.

However, there are examples of mathematical systems where features exist that can be attributed either to the bulk of the system or to the boundary\fn{The \(\theta\) angle in QCD is a case in point. One can describe that as a lack of invariance under topological gauge transformations, which can be entirely attributed to the boundary.  Equivalently, one can regard this effect as a \(PC\) violating term in the action, which is local.}, so that relegating all time symmetry violating effects to the boundary may conceivably not always work.

As long as we adhere to the quantum mechanical description of all microscopical dynamical laws, we find the CPT theorem on our way, which implies that if we combine time reversal \(T\) with parity reversal \(P\) and particle-antiparticle interchange \(C\), then this symmetry is perfect. We could well stick to our verdict that Nature's boundary conditions in the time direction suffice to explain the arrow of time.

One may observe however that  another source of time reversal asymmetry can be contemplated. As explained in previous sections, this author does not believe that ``quantum mechanics" will be the last and permanent framework for the ultimate laws of nature. If we drop it, to be replaced by some classical ideas, the need for time reversal symmetry also subsides. We could opt for an underlying theory  where information, in the classical sense, can disappear. Considering cellular automata, systems where information does get lost are much more general than the ones where information is conserved, so that switching the direction of time brings about much more dramatic changes.

How can such models lead to effective quantum theories? Does local time reversal symmetry re-emerge? We claim that, for an automaton, the possibility to generate statistical correlations that are solely based on vector space analysis, that is, vectors  evolving in Hilbert space, which lead to quantum mechanics, may be quite generic, and include models featuring information loss. 

The way to deal with information loss in this connection is very straightforward in principle, while extremely difficult in practice. The way to handle this in principle is by the introduction of \emph{information classes}: we identify the elements of an orthonormal basis of Hilbert space not with single states of the automaton, but with information classes. An information class is defined to be a class of states in an automaton that have the property that, after a finite amount of time, they all evolve to become the same state in the automaton. In principle, such classes may become extremely large, but in practice the odds of two states that resemble one another at one moment in time, to evolve into exactly the same state in the near future, might rapidly go to zero as time proceeds, so that the information classes may continue to be manageable. Formally, they might become big enough to form states that can be distinguished by only inspecting the data living on a boundary surface rather than specifying what happens in the bulk. This is what we see in the physical equations for black holes, called \emph{holography}, so that this may be seen as an indirect piece of evidence favouring underlying models with information loss.

In underlying models with information loss, the act of time reversal takes a very interesting shape: the time reverse of ontological states in Hilbert space (beables) tend to form quantum superpositions of beables in the time-reversed Hilbert space. This may perhaps explain why superpositions follow the same laws of nature as ontological states, but for the time being we just regard these generic observations to be something to keep in mind when, much like Sherlock Holmes, we attempt to figure out, in terms of models, what it might have been that actually took place, when all information we have been able to acquire, takes the shape of quantum superpositions.
\\[10pt]

\textbf{\Large Appendices}

 \appendix
 \newsecl{Superpositions and Born's probabilities}{Born}
Whenever theories with classical logic are proposed to explain quantum phenomena, 
the following questions are raised:\\
 Question 1: \emph{In Bell's experiment, a pair of particles -- call them photons -- is in an entangled state. In an ontological theory, it seems as if this pair of particles ``knows ahead of time" which superposition of states will later be chosen by Alice and by Bob for their measurements. Why does this not violate causality?}\\
 Question 2: \emph{How come that the squares of amplitudes exactly represent the probabilities for the outcomes of measurements?} (Born's rule)\\
 And question 3: \emph{What happens when a wave function collapses?} And what happens when a measurement or observation is made?
 
 These questions are all strongly related, and they can be answered together in what was advertised earlier as the Cellular Automaton Interpretation of Quantum Mechanics.\cite{GtHCA}
 
 The basic idea is that, at the tiniest distance scale that is meaningful in physics, presumably the Planck scale, around \(10^{-33}\) cm, there are laws of physics which are most efficiently formulated by not giving any reference to Hilbert space, quantum superpositions, qubits, or even action-at-a-distance. We have a \emph{cellular automaton} there, or something that resembles this very much. A cellular automaton can best be regarded as a basic computer program, where, in a massive venture of parallel computing,  digital data that are localised on some sort of grid, are being updated at the beat of an extremely fast clock. The speed of the clock may vary at some points, but these are details that we do not want to go into. Most importantly, information spreads with a limited velocity, basically the speed of light, and all this information is classical. For simplicity, we assume the system to be reversible in time, although, as was explained earlier, this might not be necessary.
 
This is clearly the kind of theory that Einstein, Bell, and many others thought they could disprove, but as we shall explain now in more detail, this is not quite the end of the story. There are various aspects of the system that need much more scrutiny, in particular the ubiquitous presence of very strong correlations at the micro-scale, which permeates to macroscopic distances, and the fact that it is fundamentally impossible to compress (to `zip') the system into a more course-grained model that reproduces all details. As soon as one tries to compress anything, uncertainties emerge that manifest themselves by looking like quantum superpositions. But I am running ahead of my arguments, let us consider the situation in a meaningful order. The more complete story is presented in Ref.\,\cite{GtHCA}.

In principle, the automaton can be in a huge number of distinct states, roughly \(2^{10^{99}}\) states in every cubic cm (a number obtained by assuming one boolean degree of freedom in a cubic Planck length). Only if we consider all of these states, the system can be seen to be deterministic. Every single one of these states is important, but, because of strong correlations, we perceive our world as if there are much less possible states, typically \(2^{10^{50}}\) in a \(\mathrm{cm}^3\) (one boolean degree of freedom in  \(1\ \mathrm{TeV}^{-3}\)). Yet compressing the system cannot be done without loosing information; a more powerful technique is required.

It so happens that a more powerful technique does exist; we call it ``vector space analysis". In mathematics, this is not new\,\fn{In physics, the most spectacular application is the solution of the 2-dimensional Ising Model, by Onsager and Kaufman\,\cite{Kauffman}. They turn the classical model into a quantum field theory that happens to be integrable.}. For instance, in group theory, it turned out to be useful to give matrix representations of elements of a group. Consider a subset of a permutation group. The elements of the set in which the permutations take place are represented as orthonormal vectors in our vector space. The dimensionality of this vector space equals the dimension (number of elements) of the set. It can be finite or infinite. This vector space is our Hilbert space. One now can use all mathematical tricks available for vectors to investigate the properties of the group. For instance, one can diagonalise the matrices. This involves orthogonal (unitary) transformations of all sorts for the vectors. 

It is now assumed that we can do the same in the set of states of the automaton. After a number of transformations, we get matrices representing the evolution that are diagonal or almost diagonal. The effective dimensionality of our Hilbert space can now be considerably reduced because large parts of it factorise. However, they do not factorise along the original dividing lines of our orthonormal set. We get different kinds of vectors, all of which are now superpositions of vectors of the original set. All of this is just mathematical manipulation; the physics is kept as it was.

In particular, the evolution law is an ontological matrix in terms of the original ontological states; an ontological unitary matrix is a matrix containing only one one and for the rest zeros in all its rows and all its columns (arbitrary phase factors are allowed, as long as each row and each column only contains one element with absolute value one, while all other matrix elements vanish). After some combination of extensive linear superpositions, our matrices will look much more generic than before.

While every one of our \(2^{10^{99}}\) states evolves into another state within time units as small as the Planck time, being of the order of \(10^{-44}\) seconds, we will find superpositions of states that evolve much slower. The effective time unit will now be the inverse of the energies of the most energetic particles in our particle accelerators. These energies are many orders of magnitude lower than the Planck energy, so indeed, we have a much smaller Hilbert space than the original one. What is known about physics today is the evolution laws of this tiny subspace of Hilbert space. Since the time dependence is much slower here, we can write the evolution law in terms of a hermitian hamiltonian: the Schr\"odinger equation. We only postulate determinism in the original cellular automaton model with its humongous number of states, not in the effective, reduced model that is called physics today. Can this system violate the Bell/CSHS inequalities?

First we need to specify how an observation is made, in terms of the states of the original automaton. Suppose we want to establish the presence of a planet. In the interior of the planet, atoms and molecules are densely packed, so that the world in there looks quite different from the vacuum state. We now assume that the vacuum state is represented by states in the automaton that show different statistical abundances and correlations than the states that represent densely  packed atoms and molecules. Locally, the statistical differences between these states may be minute; our ability to distinguish the vacuum state from the rocky material may be far from perfect; say that, inside a small volume of a \(\textrm{mm}^3\), a given state has a likelihood of \((1-\e)\,/\,(1+\e)\) of being a vacuum rather than a rock. For the whole planet, we have to raise this number to a power equal to the volume of the planet measured in  \(\textrm{mm}^3\). Thus one finds almost with certainty that there is a planet rather than a vacuum in that neighbourhood.

The planet is a classical object. What we just found is that such classical objects are bound to be sufficiently well identified and characterised in terms of the original states of the automaton. Let us assume that this holds for all objects that we normally call ``classical", not necessarily as large as planets. When we do a measurement or make an observation, we must be looking at a large subset of the classical states of the automaton.
	
Now consider a quantum experiment. We can't use the entire Hilbert space, because it contains far too many states. So we use the strongly reduced subspace of Hilbert space that represents only low-energy particles. All these states are superpositions of cellular automaton states. Specifying our initial state \(|\psi\ket_\ini\) as well as we can, we still represent it as a superposition of ontological states \(|\ont\ket_i\):
	\be|\psi\ket_\ini=\sum_i\a_i|\ont,\,\ini\ket_i\ ;\quad\sum_i|\a_i|^2=1\ .\eel{psiini}
At this point we merely need to \emph{define} that \(|\a_i|^2\) represents the probability that the ontological state \(|\ont\ket_i\) is our initial state. From the mathematics of linear representation theory, it would be hard to deduce any other link between probabilities and amplitudes than that one. In any case, in what follows, we shall see that what holds for the initial state will continue to hold for all states arrived at in later times.

So let us consider the evolution of this state. Our mathematical procedures for the decompositions of our state vectors never affected the physical evolution law for the ontological states. This means that, \emph{as long as we use linear Schr\"odinger equations}, also at later times, relation \eqn{psiini} continues to hold, up to the final state:
	\be|\psi\ket_\fin=\sum_i\a_i|\ont,\,\fin\ket_i\ ;\quad\sum_i|\a_i|^2=1\ .\eel{psifin}
Note that the basis of states will have changed, but the superposition coefficients \(\a_i\) have stayed exactly the same, and hence also the probabilities stayed the same\fn{We often get the question whether taking the absolute squares of \(\a_i\) as being the probabilities, doesn't change everything. The answer is no, because the coefficients do not change at all during the entire evolution, as long as we stay in the ontological basis.}. And now consider the measurement. We compare the final superimposed state with the ontological states the system should end up in. They are again the ontological states \(|\ont,\,\fin\ket_i\) of Eq. \eqn{psifin}. Now the \(\a_i\) are finally recognised as representing the probabilities for the final state. Born's probability rule is the simple consequence of the mathematical representation theory. The answer to the question where Born's probability rule comes from is that, if we put it in for the initial state, Born's rule stays the same during the entire evolution. 

Note now that, if we started with one single ontological state \(|\ont,\ini\ket_1\), then the final state will \emph{automatically} also be a single ontological state \(|\ont,\fin\ket_1\). This continues to be true if we use the Schr\"odinger equation to describe the evolution. Consequently, the Schr\"odinger equation will \emph{automatically} cause the final state to collapse into a single ontological state, if the initial state was a single ontological state. The reason why this appears not to happen in ordinary quantum mechanics is that we do not use the full Schr\"odinger equation for all states, but only for the lower energy states where the equation is known, and we idealised the initial state, involuntarily replacing the ontological initial state by a superposition, hence a probabilistic distribution of initial ontological states. 

It is often claimed that quantum probabilities should be seen as fundamentally different from the classical uncertainties that are due to lack of knowledge of the initial state; in our approach however, the quantum probabilities are there for exactly the same reasons as in classical theories.
	
	Now consider the EPR / Bell experiment. We do not explicitly construct a microscopic, classical model for all Standard Model interactions. Although general strategies for such a construction have been proposed, it is still too difficult to reproduce all symmetries of Nature. We do however claim that any contradiction with the Bell/CHSH inequalities has disappeared. 
	
	When Alice and Bob perform their observation, they cannot select a superposition of photon states, but only one ontological photon. The outcome of Alice's  measurement is always an  ontological state of the form \(|a,\,A\ket_\ont\), where \(a\) is the setting chosen, given by an angle, and \(A=\pm 1\) is her finding. Together with Bob's finding, the final, classical state is \(|a,\,A,\ b,\,B\ket_\ont\). In our model, the calculation gives a superposition, 
\beq\psi\ket_\fin&\ =\ \a_1|a,+,\ b,+\ket_\ont&+\ \ &\a_2|a,+,\ b,-\ket_\ont\\
			&\ +\ \a_3|a,-,\ b,+\ket_\ont&+\ \ &\a_4|a,-,\ b,-\ket_\ont\ ,\eeql{abfinal}
	
	The observed outcome is never a final state of the form \eqn{psifin} or \eqn{abfinal}, but \emph{always} one specific ontological state, \(|\ont,\,\fin\ket_{1}\).  The \emph{model calculation} gives an entangled superposition of the ontological state \(|a,b\ket\) combined (multiplied) with a superposition of the four states \(|+,+\ket,\ |+,-\ket,\ |-,+\ket,\ \) and \(|-,-\ket\).
	
	If we modify the initial state, the calculated final state will be a different entangled superposition, but the ontological state will be in the basis of the angles \(a,b\) and the measurements \(A\) and \(B\).
Modifying the initial ontological state will always lead to a single final ontological state, \emph{never} a superposition, since the coefficients \(\a_i\) never change.

	What was misleading in Bell's exposition of the experiment is that he thought that a modification of the settings \(a\) and \(b\) would lead to a different superposition of the measurements \(A=\pm\) and \(B=\pm\). In our vector representation, any modification of \(a\) and \(b\), regardless how tiny,  requires a modification of the initial ontological state. The new ontological state will be orthogonal, hence totally unrelated to the previous one, so that the two photons emitted by the source cannot be related to the photons emitted previously. Thus, the idea that one can modify the settings \((a,b)\) without modifying the polarisation of the entangled photons emitted by the source, is an illusion.
	
	One can also say that the settings \(a\) and \(b\) emerge to be \emph{entangled} with the polarised photons. As soon as the settings are fixed, the photons will only be in a single ontological state. I won't push this description too much, because at the end we should have just a single setting and a single ontological photon state.
	
	The most important difference between our presentation and the usual treatment of Bell's observations is that the observers Alice and Bob, together with the settings \(a\) and \(b\) chosen by them,  \emph{are parts of the physical system}.  Any modification of the settings \((a,b)\), whether done out of ``free will" or otherwise, will require a different initial ontological state.

 \newsecl{The GHZ paradox and the 6-bit universe}{6bit}
 There are many newer versions, generalisations and refinements of the original Gedanken experiments considered by EPR and Bell. Sometimes, the paradoxes concern not only probabilities, but even certainties where clashes with ``classical" physics are seen to occur, but they all have in common that one or more observers choose between two or more different settings that measure properties of quantum objects, whose operators do not commute. 
 
 An interesting case, where the magic mystery seems to reach new heights, is the GHZ paradox. We briefly recapitulate the setup, which is explained in more detail in te literature\,\cite{GHZ,mermin}. 
 
 A source is constructed such that it emits three entangled particles, each having two possible spin states, \(\pm 1\). The quantum state produced is
 	\be \psi=\fract 1{\sqrt 2}\big(\,|+,+,+\ket - |-,-,-\ket\,\big)\ . \eel{GHZini}
The operators to be considered are \(\s_{x,\,y}^{a,\,b,\,c}\) where \(a,\,b\) and \(c\) refer to the three particles \(a, \,b\) and \(c\), and
	\be \s_x^a|\pm,\, \cdots\ket=|\mp,\,\cdots\ket\ ,\quad \s_y^a|\pm,\, \cdots\ket=
	\pm i|\mp,\,\cdots\ket\ ,\eel{threesigmas}
 while \(\s_{x,\,y}^b\) and \(\s_{x,\,y}^c\) act similarly on particle \(b\), and on \(c\), respectively. It is not difficult to derive that these operators obey
 	\beq XXX \equiv&\ \s_x^a\,\s_x^b\,\s_x^c&=&\ -1 \ ,\\
		XYY \equiv&\ \s_x^a\,\s_y^b\,\s_y^c&=&\ 1\ ,\\
		YXY \equiv&\ \s_y^a\,\s_x^b\,\s_y^c&=&\ 1\ ,\\
		YYX \equiv&\ \s_y^a\,\s_y^b\,\s_x^c&=&\ 1\ . \eeql{sigmaprods}
The three Pauli matrices \(\s_i\) acting on the same particle, anti-commute, \(\s^a_x,\,\s^a_y=-\s^a_y\,\s^a_x\), while two Pauli matrices acting on different particles commute. Thus one derives that if we permute two pairs of \(\s\) operators in Eq.~\eqn{sigmaprods}, two minus signs emerge, which enables us to derive easily that all four operators in Eq.~\eqn{sigmaprods} commute with one another. Therefore, all operators in Eq.~\eqn{sigmaprods} can be measured simultaneously, and the result always obeys \eqn{sigmaprods}.	
		
		Now, the three particles are sent to three different observers, who sit in three different, sealed rooms. Each observer decides, ``at his free will", to choose to measure either \(X=\s_x\) or \(Y=\s_y\). The observers cannot communicate with each other, so they do not know what the others choose. They just meticulously write down whether they measure \(X\) or \(Y\), and what their outcome is, \(+1\) or \(-1\). After having done a long series of measurements, they come out of their rooms, and compare notes.
		
		All observers, on average, found as many pluses as minuses, because the expectation values of \(X=\s_x\) and \(Y=\s_y\) are zero. Also, there is no pair correlation, since for every pair, the expectation values for \(XX\), \(XY\), \(YX\) and \(YY\) are also zero. But the three observations are correlated: the three-point correlations, given in Eq.~\eqn{sigmaprods}, are very strong.
		
		Moreover, they seem to contradict classical logic. The list of observations will obey \eqn{sigmaprods}. But at every run, one might have asked: what would this observer have found if (s)he chose the other setting, or more generally, given a particle entering his room, and (s)he measures either \(X\) or \(Y\), what would the outcome in either case have been? So we add to the list of observations, at each run, all possible answers: \(XXX,\,XXY,\, \cdots,\,YYY\). Now take the last three equations of \eqn{sigmaprods}. Take their product. Since every \(Y\) occurs exactly twice in the product, the \(Y\)s together always contribute \(+1\) in the product. What is left is the three \(X\)s. So we get \(XXX=+1\). But this is wrong, it violates the first equation of Eq.~\eqn{sigmaprods}. One must conclude that the three entangled particles know, ahead of time, whether their observers will have chosen \(X\) or \(Y\). Apparently, the observations that were not actually made, do not have well-defined values for \(X\) or \(Y\) at all. These are called \emph{counter factual}. Quantum mechanics forbids counter factual observations. How can this happen in a cellular automaton?
		
		In this case, vector space analysis suggests that a simple model can be constructed of the entire universe. There are just 6 binary dynamical variables in this universe. A priori, this universe could have started choosing any of \(2^6=64\) distinct initial states.
		
		Like our real universe, this model universe may have started out with a big bang. At that moment, not all possible states have been realised. Only 48 of the 64 initial states were allowed. During a period of chaos, the 48 states may have been scrambled many times, but there are 16 states that cannot be realised at any time. This is how the laws of nature for the model universe are programmed.
		
		At the beginning of the experiment, three particles are selected. These are three of the 6 bits. All of them can be \(+1\) or \(-1\). Now we have the three observers, \(A,\,B,\) and \(C\). Each of them has to decide whether to choose \(X\) or \(Y\). They each grab the one bit they can find in their rooms. That bit represents their free will. It can be anything, but its properties are determined by laws of nature. Each observer knows that the probability for this bit to be \(+1\) or \(-1\) will be equal, so the observers will be convinced they are acting out of free will. There are \(2^3=8\) possible terms in the sequence \(XXX,\,YXX,\,\cdots,\,YYY\). In 4 of these (where the number of \(Y\)s is even), there is a constraint: only 4 of the \(2^3=8\) possible answers are allowed. Therefore, \(4\times 4=16\) outcomes are forbidden. This is what the laws of nature tell you here: of all ontological states, 16 are forbidden.
		
		Thus, we claim that classical laws of nature in the 6 bit universe can perfectly well reproduce the GHZ ``miracle", but we must accept that the observer's free will is controlled by laws of nature as much as all other phenomena.
		
		Of course, quantum physicists object that this is unfair: ``you have used `retro-causality' to establish your laws of nature". Well, the view presented in the main body of this paper is, that the laws of nature  are usually time-reversal invariant, and this means that if a complete state of the universe is known at present, it also causes limitations to the allowed states in the past, and that is where our constraints come from. We simply cannot expect `perfect' free will in our universe. Maybe you think this is `conspiracy'. So be it, but the laws of nature in our approach are foremost classical.\\[10pt]
		
\noindent\textbf{\large Acknowledgment}\\

{The author thanks T.~Maudlin, P.W. Morgan, T.~Myers, T~Norsen, 
and many others, for extensive discussion on these and related issues.}

\end{document}